\documentclass[twoside,twocolumn,9pt]{article}
\usepackage{extsizes}
\usepackage[super,sort&compress,comma]{natbib} 
\usepackage[version=3]{mhchem}
\usepackage[left=1.5cm, right=1.5cm, top=1.785cm, bottom=2.0cm]{geometry}
\usepackage{balance}
\usepackage{mathptmx}
\usepackage{sectsty}
\usepackage{graphicx} 
\usepackage{lastpage}
\usepackage[format=plain,justification=justified,singlelinecheck=false,font={stretch=1.125,small,sf},labelfont=bf,labelsep=space]{caption}
\usepackage{float}
\usepackage{fancyhdr}
\usepackage{fnpos}
\usepackage[english]{babel}
\addto{\captionsenglish}{%
  
}
\usepackage{array}
\usepackage{droidsans}
\usepackage{charter}
\usepackage[T1]{fontenc}
\usepackage[usenames,dvipsnames]{xcolor}
\usepackage{setspace}
\usepackage[compact]{titlesec}
\usepackage{hyperref}
\usepackage{epstopdf}%This line makes .eps figures into .pdf - please comment out if not required.

\definecolor{cream}{RGB}{222,217,201}

\begin{document}

\pagestyle{fancy}
\thispagestyle{plain}
\fancypagestyle{plain}{
%%%HEADER%%%
\renewcommand{\headrulewidth}{0pt}
}
%%%END OF HEADER%%%

%%%PAGE SETUP - Please do not change any commands within this section%%%
\makeFNbottom
\makeatletter
\renewcommand\LARGE{\@setfontsize\LARGE{15pt}{17}}
\renewcommand\Large{\@setfontsize\Large{12pt}{14}}
\renewcommand\large{\@setfontsize\large{10pt}{12}}
\renewcommand\footnotesize{\@setfontsize\footnotesize{7pt}{10}}
\makeatother

\renewcommand{\thefootnote}{\fnsymbol{footnote}}
\renewcommand\footnoterule{\vspace*{1pt}% 
\color{cream}\hrule width 3.5in height 0.4pt \color{black}\vspace*{5pt}} 
\setcounter{secnumdepth}{5}

\makeatletter 
\renewcommand\@biblabel[1]{#1}            
\renewcommand\@makefntext[1]% 
{\noindent\makebox[0pt][r]{\@thefnmark\,}#1}
\makeatother 
\renewcommand{\figurename}{\small{Fig.}~}
\sectionfont{\sffamily\Large}
\subsectionfont{\normalsize}
\subsubsectionfont{\bf}
\setstretch{1.125} %In particular, please do not alter this line.
\setlength{\skip\footins}{0.8cm}
\setlength{\footnotesep}{0.25cm}
\setlength{\jot}{10pt}
\titlespacing*{\section}{0pt}{4pt}{4pt}
\titlespacing*{\subsection}{0pt}{15pt}{1pt}
%%%END OF PAGE SETUP%%%

%%%FOOTER%%%
\fancyfoot{}
\fancyfoot[LO,RE]{\vspace{-7.1pt}\includegraphics[height=9pt]{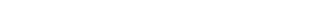}}
\fancyfoot[CO]{\vspace{-7.1pt}\hspace{13.2cm}\includegraphics{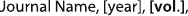}}
\fancyfoot[CE]{\vspace{-7.2pt}\hspace{-14.2cm}\includegraphics{head_foot/RF}}
\fancyfoot[RO]{\footnotesize{\sffamily{1--\pageref{LastPage} ~\textbar  \hspace{2pt}\thepage}}}
\fancyfoot[LE]{\footnotesize{\sffamily{\thepage~\textbar\hspace{3.45cm} 1--\pageref{LastPage}}}}
\fancyhead{}
\renewcommand{\headrulewidth}{0pt} 
\renewcommand{\footrulewidth}{0pt}
\setlength{\arrayrulewidth}{1pt}
\setlength{\columnsep}{6.5mm}
\setlength\bibsep{1pt}
%%%END OF FOOTER%%%

%%%FIGURE SETUP - please do not change any commands within this section%%%
\makeatletter 
\newlength{\figrulesep} 
\setlength{\figrulesep}{0.5\textfloatsep} 

\newcommand{\topfigrule}{\vspace*{-1pt}% 
\noindent{\color{cream}\rule[-\figrulesep]{\columnwidth}{1.5pt}} }

\newcommand{\botfigrule}{\vspace*{-2pt}% 
\noindent{\color{cream}\rule[\figrulesep]{\columnwidth}{1.5pt}} }

\newcommand{\dblfigrule}{\vspace*{-1pt}% 
\noindent{\color{cream}\rule[-\figrulesep]{\textwidth}{1.5pt}} }

\makeatother
%%%END OF FIGURE SETUP%%%

%%%TITLE, AUTHORS AND ABSTRACT%%%

\LARGE{\textbf{A Reaction-Diffusion-Chemotaxis Model for Human Population Dynamics over Fractal Terrains$^\dag$}} \\%Article title goes here instead of the text "This is the title"
\large{Benjamin M. Alessio,\textit{$^{a,b}$} and Ankur Gupta$^{\ast}$\textit{$^{b}$}} \\%Author names go here instead of "Full name", etc.

\normalsize{Advection of entities induced by gradients in attractant concentration fields is observed via diffusiophoresis in colloids and via chemotaxis in microorganisms. Mathematically, both diffusiophoresis and chemotaxis follow similar mathematical descriptions and display a variety of interesting behaviors that are not observed through other transport mechanisms. However, the application of such a mathematical framework has largely been restricted to soft matter research. In this article, we argue that this framework is more general and can be expanded to study human population dynamics. We assert that human  populations also migrate chemotactically, but by sensing concentrations gradients in attractants such as resource availability, social connections, and safety indices. Therefore, we extend the Fisher-KPP reaction-diffusion model, foundational to human population dynamics, to incorporate chemotactic advection. Furthermore, we introduce a fractal terrain to better mimic the human dispersal phenomena. Simulations demonstrate that, by including chemotaxis of a population toward attractants which are dispersed heterogenously over fractal terrains, population hotspots can appear from from initially uniformly dispersed states whereas Fisher-KPP without chemotaxis predicts a persistent tendency toward population uniformity. Varying the chemotactic migration yields fine control over inter- or intra-population segregation and thus the population growth rates may be substantially altered by considering the population-attractant coupling. This framework may be useful for characterizing historical population separations, and furthermore is particularly pertinent for predicting emergence of new population hotspots as climate change is expected to cause large-scale human displacement, which may be dictated by chemotactic movement of humans due to evolving concentration gradients in safety indices.} \\

%%%END OF TITLE, AUTHORS AND ABSTRACT%%%

%%%FONT SETUP - please do not change any commands within this section
\renewcommand*\rmdefault{bch}\normalfont\upshape
\rmfamily
\section*{}
\vspace{-1cm}

%%%FOOTNOTES%%%

\footnotetext{\textit{$^{a}$Department of Mechanical Engineering; Stanford University, CA}}
\footnotetext{\textit{$^{b}$Department of Chemical and Biological Engineering; University of Colorado Boulder, CO}}
\footnotetext{\textit{$^*$E-mail: Ankur.Gupta\@Colorado.Edu}}

%Please use \dag to cite the ESI in the main text of the article.
%If you article does not have ESI please remove the the \dag symbol from the title and the footnotetext below.

%%%END OF FOOTNOTES%%%

%%%MAIN TEXT%%%%
\section{Introduction}
Descriptions of human populations dynamics, whether for assessing modern problems, inquiring about our distant past, or making predictions of the future, rely heavily on mathematical modeling because data is scarce~\cite{steele2009human}. Mathematical modeling of the dispersal of human populations is based largely on reaction-diffusion frameworks with population growth and death, originating from the seminal works of Fisher~\cite{fisher1937wave}, Kolmogorov, Petrovski, and Piskunov~\cite{kolmogorov1937study}, and Skellam~\cite{skellam1951random} (Fisher-KPP framework). Together, these effects can be responsible for a travelling population wave whose speed increases with either the coefficient of diffusion or the growth rate. This salient feature of the Fisher-KPP model has been cited to propose the random walk as a key mechanism for out-of-Africa theory~\cite{young1995simulating}, the Neolithic transition~\cite{fort1999time}, dynamical evolution~\cite{deforet2019evolution}, colonization~\cite{hamilton2007spatial}, spread~\cite{young1995simulating} and evolution~\cite{deforet2019evolution}. Efforts to improve quantitative capabilities of models which build off of the Fisher-KPP framework ~\cite{steele2009human} focus on including  advection~\cite{cohen1992nonlinearity, davison2006role, gu2017stationary, timmermann2016late}. These efforts either directly specify a non-dynamic velocity vector field~\cite{timmermann2016late} or assume that advective velocity is dependent on spatially heterogeneous carrying capacity~\cite{cohen1992nonlinearity,davison2006role,gu2017stationary}.  However, these velocity forms are included in an ad-hoc manner, which leaves open to question the true source of advection.

This is in stark contrast to the literature on soft matter physics, which, although originally based on the same Fisher-KPP framework, has employed first-principles reasoning for advective motion~\cite{woodward1995spatio,brenner1998physical,ryzhik2008traveling,marel2014flow,hamel2020propagation}. Specifically, advection under the umbrella of \textit{diffusiophoresis}, i.e., transport due to the concentration gradient of an attractant, is central to several fundamental processes. For instance, in colloidal suspensions, diffusiophoresis induces a velocity to particles proportional to chemical concentration gradients of a diffusing solute~\cite{shim2022diffusiophoresis, anderson1984diffusiophoresis, ganguly2023diffusiophoresis}. The spatial gradients of the solutes have many possible origins~\cite{velegol2016origins, jarvey2023asymmetric}, such as the diffusion of an initially concentrated source~\cite{ault2017diffusiophoresis}, reaction-diffusion instabilities (a.k.a. Turing Patterns)~\cite{alessio2023diffusiophoresis}, or asymmetric reactions on the colloid surface~\cite{brady2011particle, ganguly2024unified} or asymmetry in geometry~\cite{raj2023motion, ganguly2023going}. Diffusiophoresis has been shown in experiments to induce flow in regions which are inaccessible to pressure-driven flows, such as fibrous materials~\cite{shin2018cleaning} or dead-end pores~\cite{alessio2021diffusiophoresis}. Furthermore, a key feature of diffusiophoresis is the
focusing of colloids~\cite{shi2016diffusiophoretic, raj2023two}. In quasi-one-dimensional systems, this sharpening can be quantified when accounting for confining boundary effects which can impose advective fluxes that appear macroscopically as diffusive fluxes~\cite{alessio2022diffusioosmosis,ding2023shear}. In active matter systems, this sharpening manifests as clustering or phase separations~\cite{cates2015motility} and specific interactions give rise to lifelike properties~\cite{zeravcic2017colloquium}.

In addition to colloids, concentration gradients can direct the motion of organisms. This is referred to as \textit{chemotaxis} when an organism moves along concentration gradient of a chemoattractant~\cite{murray2002reaction}. The organism senses the direction of increasing chemical concentration and decides to propel itself accordingly. Chemotaxis, which is mathematically identical to diffusiophoresis~\cite{chu2021macrotransport, alessio2023diffusiophoresis}, is macroscopically an advective process which arises from a microscopic random walk that is biased in the direction of the chemical gradient~\cite{keller1971model}. It is typical in chemotactic systems for the chemical concentration to depend on the organism concentration, as in communication and directed motion \textit{via} pheromones. Keller and Segel's seminal model of slime mold~\cite{keller1970initiation} includes this feedback mechanism where amoebae produce two chemical solutes which react and form complicated spatial gradients that then guide the amoebae. By including advection based on chemical gradients as opposed to purely reaction-diffusion analysis \cite{turing1952chemical, prigogine1967symmetry}, Keller and Segel ignited decades of research on the role of chemotaxis in biophysics~\cite{horstmann20041970,murray2002reaction, zhao2023chemotactic, hardt2020electric}. 

As is clear from the above discussion, diffusiophoresis and chemotaxis play a pivotal role in understanding colloidal physics and chemotactic systems. The  central argument of this article is that much like colloids and microorganisms, humans also advect along gradients of attractant landscapes. However, unlike colloids and microorganisms that sense gradients in chemical species, humans sense gradients in multiple attractants such as economic opportunities,  social connections and safety indices, among other factors, and advect by weighing the gradients appropriately. However, there are some additional complications that arise for human population dynamics. The resource availability and population growth in humans is likely going to be spatially heterogenous. Distinctions between the diffusive behavior of random walkers on uniform and on fractal terrains have been well-studied in the physics literature~\cite{campos2004description} and used to explain anomalous migration rates of ancient populations~\cite{hamilton2007spatial} and animal population dynamics~\cite{johnson1992animal}. Therefore, in addition to gradient-based advection, we must also consider that the movement of humans occurs over a fractal terrain.

There have been some attempts to include these effects for human population dynamics. For instance, previous studies~\cite{davison2006role,timmermann2016late} have recognized the role of terrain (specifically waterways) and accordingly implemented advection as an ad-hoc fitting parameter for when a random walk model fails. Furthermore, Cohen~\cite{cohen1992nonlinearity} proposed a chemotaxis-like advection of human populations due to gradients in carrying capacity as a mechanism for subsidence boundary formation. We also note that a similar model was explored by Gu et al~\cite{gu2017stationary} in the context of urban crime patterns, where a criminal population drifts toward regions of increasing house attractiveness. To our knowledge, no study has extended Cohen's theory to demonstrate in general the ability a heterogeneous terrain to incite population clustering. By incorporating a heterogeneous, fractal terrain-dependent chemotactic advection into a Fisher-KPP model, we aim to reveal a physical mechanism through which human populations may naturally become highly heterogeneous. This is particularly relevant, as the evolution of population hotspots is one of the expected outcomes of climate change. While the purpose of this article is not to claim that the model outlined below can make such a prediction, it does suggest essential features of a mathematical model that might be necessary for such predictions in the future.

\begin{figure}[h]
\includegraphics[width=0.45\textwidth]{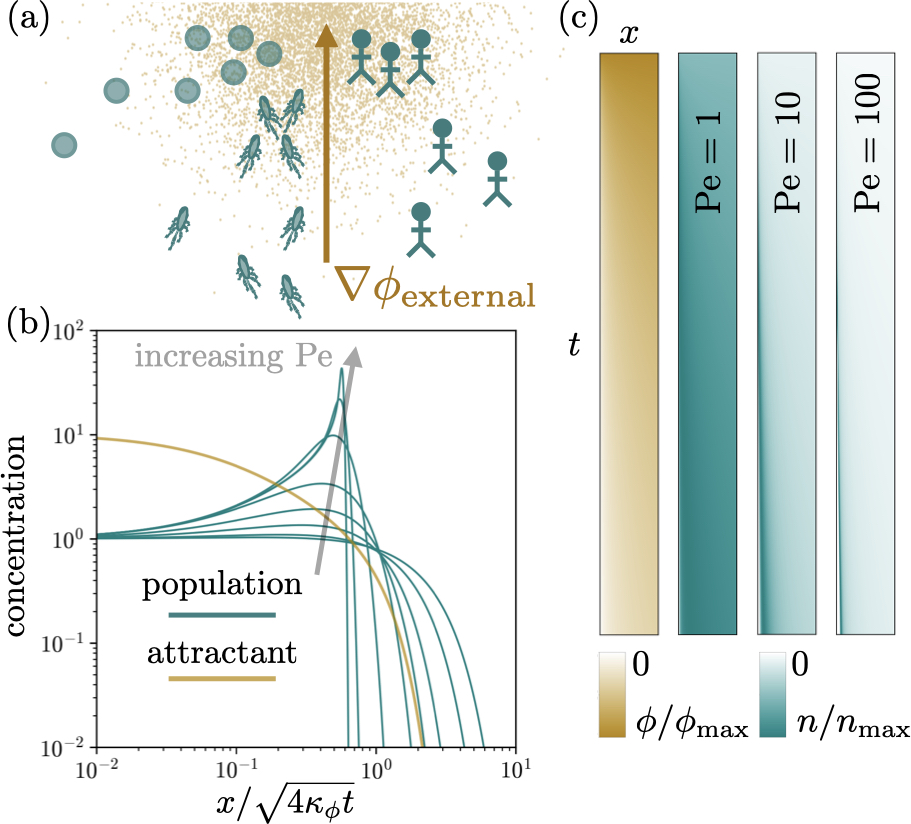}
\caption{\textbf{Chemotactic focusing through a diffusive attractant gradient}. A key feature of chemotaxis or diffusiophoresis is the spatial focusing of concentrations due to migration along the gradient of an attractant. Owing to the nonlinearity of the migration term, the gradients in the chemotactic concentration in general develop a length scale which is distinct from that of the attractant, which itself may be limited to diffuse length scales. The strength of the attraction and the diffusivity of the population determine the Péclet number which controls the extent to which the population focuses. For a large Péclet number, the chemotactic population may become orders of magnitude more intensely concentrated than the attractant. $(a)$ Schematic of a population which sharpens due to an externally applied purely diffusive attractant gradient. $(b)$ Sharpening of a population with density $n$ vs. similarity variable $\eta$ increases with $\text{Pe} = \frac{\chi}{\kappa_N}$. $(c)$ The density profiles (teal), normalized by their maximum value, are shown in  from $(b)$ are shown in space (horizontal) and time (vertical) alongside the diffusive attractant gradient (gold). $\beta=0.1$ is used for all calculations. }
\label{fig:sharpening}
\end{figure}

\section{Mathematical Model}
We solve a system of partial differential equations (PDEs) for population(s) density $n$, described by a Keller-Segel style chemotactic reaction-diffusion equation (\ref{eqn:kellersegel}) and attractant(s) $\phi$ described by a reaction-diffusion equation (\ref{eqn:phi}), for diffusivities $\kappa_n$ and $\kappa_\phi$, growth rates $r_n$ and $r_\phi$, and chemotactic mobility $\chi$ (all of which may in general vary spatiotemporally):
\begin{subequations}
\label{Eq: geqn}
\label{eqn:governingeqs}
\begin{eqnarray}\label{eqn:kellersegel}
\frac{\partial n}{\partial t} = \nabla\boldsymbol\cdot \left(\kappa_n\nabla n - v n  \right) + r_n \\
v = \chi\nabla\phi \textrm{ or } v= \chi \nabla \log \phi, \\
\label{eqn:phi}
\frac{\partial \phi}{\partial t} = \nabla\boldsymbol\cdot\left(\kappa_\phi\nabla\phi\right) + r_\phi.
\end{eqnarray}
\end{subequations}
As discussed above, in reality, there are going to multiple attractant fields, in which case these equations may be expanded to $\left[\phi_1, \phi_2, ...\right]$. The choice of a reaction-diffusion equation to describe $\phi$ is not central to our argument but provides a good starting point for capturing key characteristics of physical fields that may serve to attract the population $n$, such as economic potential and intellectual property~\cite{buera2020global,stoneman1985technological}, or prey animal populations~\cite{johnson1992animal}. Furthermore, we note that the reaction-diffusion equation alone, if sufficiently nonlinear, can permit heterogeneity such as pattern formation in soft matter systems~\cite{turing1952chemical,alessio2023diffusiophoresis}. There may also be multiple populations $\left[ n_1, n_2, .. \right]$ which will have multiple mobility coefficients $\chi_{ij}$ for $i^{\textrm{th}}$ population and $j^{\textrm{th}}$ attractant. To retain simplicity, we avoid multiple species and attractants. In our system, we connect a spatially varying terrain $h(\boldsymbol x)$ to the dynamics through simple elevation-dependent growth rates
\begin{subequations}
\label{Eq: rxns}
\begin{equation}
    \label{eqn:popgrowth}
    r_n(n) = \underbrace{ r_{n0} n\left(1-\frac{n}{n_\text{max}}\right) }_\text {Fisher-KPP} \underbrace{\exp\left[-\left(\frac{h(\boldsymbol x)-h_{0n}}{\lambda_n}\right)^2\right] }_\text{terrain},
\end{equation}
\begin{equation}
    \label{eqn:attgrowth}
    r_\phi(\phi) = \underbrace { r_{0\phi} \phi\left(1-\frac{\phi}{\phi_\text{max}}\right) }_\text{Fisher-KPP} \underbrace{\exp\left[-\left(\frac{h(\boldsymbol x)-h_{0\phi}}{\lambda_\phi}\right)^2\right]}_\text{terrain} + \underbrace{ r_{\textrm{external}} }_\text{external},
\end{equation}
\end{subequations}
where $r_{0n}$ and $r_{0\phi}$ are prefactors, $n_\text{max}$ and $\phi_\text{max}$ are the maximum concentrations beyond which the growth rate becomes negative, and the populations only grows within Gaussian envelopes of elevation centered at $h_{0n},~h_{0\phi}$ and widths $\lambda_n,~\lambda_\phi$. These parameters $h_0$ and $\lambda$ describe in an idealized way the spatial preferences for a population (or attractant) regarding where it will grow most quickly, i.e. $h_0$ may refer an optimal elevation while $\lambda$ may describe how far, under no external influences, the population would venture away from $h_0$. Given the breadth of our message, we do not consider a complete exploration of the parameters in equations~\ref{Eq: geqn} and~\ref{Eq: rxns} appropriate so we select exemplary parameters. For the attractant field,  $r_{\textrm{external}}$ denotes the field beyond the Fisher-KPP's second-order reaction, which could be included, but is not the focus of our work.

\section{Solution Procedure}
We simulate this unsteady system in two spatial dimensions using Basilisk C~\cite{popinet2015quadtree}, an open source software for solving PDEs. The system of reaction-diffusion-advection equations is solved using their diffusion module with a time-implicit scheme, and their advection module which uses a second-order unsplit upwind scheme. The automatic mesh refinement feature of Basilisk enables resolution of the multiple spatial scales inherent to our problem extending from the domain size to the diffusive scale below which the fractal nature is not relevant. Two kinds of terrains $h(\boldsymbol x)$ are used: a simple terrain is generated using a combination of randomly placed Gaussian elevation profiles with length scales on the order of 10\% of the domain size, and a fractal terrain is generated using the midpoint displacement algorithm~\cite{norros1999simulation} and interpolated linearly onto the adapting grid~\footnote{See basilisk.fr/sandbox/benalessio for code implementations.}. For simplicity we select all parameters to be uniform in space except for $h(\boldsymbol x)$. We impose uniform initial conditions beneath the carrying capacity for both $n$ and $\phi$, and always use no-flux conditions for $\phi$ at all four boundaries. For traveling wave simulations, $n$ is initialized to zero while the top and bottom boundaries are no-flux, the left is $n=n_0>0$, and the right is $n=0$; for steady-state simulations no-flux is imposed on all boundaries and $n$ is initialized to $n_0>0$. We note that the specific parameters selected in the simulations are not indicative of any physical model and are chosen  to demonstrate the model features of interest.

\section{Chemotactic focusing on a uniform terrain with no reactions} We initially start with a straightforward model where we do not include reactions or the effect of heteregenous terrain. Even with this simple model, there are some distinctive features that the proposed model yields that are not possible from the Fisher-KPP framework. One of these features is a shock-wave-like spatial focusing, where an initially spread-out population will become highly focused when an attractant gradient is introduced. This feature can be easily illustrated through a semi-infinite one-dimensional model with an external diffusive gradient (figure~\ref{fig:sharpening}(a)). Following the framework of Ault et al.~\cite{ault2017diffusiophoresis} for colloids, we simplify equation~\eqref{Eq: geqn} to $\frac{\partial n}{\partial t} = \kappa_n \frac{\partial^2 n} {\partial x^2 } - \frac{\partial}{\partial x} \left(  v  n \right)$ and $\frac{\partial \phi}{\partial t} = \kappa_\phi\frac{\partial^2\phi}{\partial x^2}$, where $\kappa_n$ is the diffusivity of the population, $v=\chi \frac{\partial\ln\phi}{\partial x}$ is the chemotactic velocity (log-sensing relationship), $\chi$ is the chemotactic mobility, and $\kappa_\phi$ is the diffusivity of the attractant.

We assume attractant conditions $\phi(0,t)=\beta$, $\phi(\infty,t)=1$, and $\phi(x,0)=1$ which correspond to diffusion from a constant source at $x=0$ into a semi-infinite reservoir. We introduce the similarity variable $\eta=\frac{x}{\sqrt{4 \kappa_\phi t}}$ to obtain the analytical form of the attractant $\phi(\eta) = \beta + (1-\beta)\text{erf}(\eta)$ and the linear ordinary differential equation for the population $\frac{\kappa_n}{\kappa_\phi}n'' + 2\eta n' - \frac{\chi}{\kappa_\phi}\left((\ln\phi)'n' + n(\ln\phi)''\right)=0$. Finally assuming an initially homogeneous population distribution, we simulate this system of equations. The results are summarized in figure~\ref{fig:sharpening}(b)-(c) with $\beta=0.1$. 

We set the diffusivities such that $\frac{\kappa_n}{\kappa_\phi} \ll 1 $ since the attractant typically diffuse faster than the population. The Péclet number, $\text{Pe}=\frac{\chi}{\kappa_n}$, is a useful measure of the strength of advection compared to the strength of diffusivity for the population. In figure~\ref{fig:sharpening}(b), population is plotted versus the similarity variable for three orders of magnitude of Pe to demonstrate the intense sharpening effect of advection. In figure~\ref{fig:sharpening}(c), the attractant and population (normalized by maximum value) is plotted in space and time to demonstrate the traveling shock wave behavior of a diffusiophoretic population advecting and diffusing along an external diffusive gradient. In the context of human population dynamics, this feature demonstrates how relatively shallow attractant gradients such as food availability or economic opportunity can lead to intense increases in population density. These population density increases may be desirable features for modeling the formation of urban hotspots or tendencies toward overpopulation from an initially dispersed human population.

\par{} We acknowledge that simulation of hotspots do have some limitations. First, the equations do not account for the upper limit of population density, and extrapolate dilute theory of diffusion to a dense system. However, this effect can be incorporated in the equations by adding an entropic penalty via the so-called Bikerman correction~\cite{kilic2007steric, gupta2018electrical}. We anticipate that by including the Bikerman correction, the extreme segregation will have an upper limit dictated by the maximum population density allowable. Second, the mobility of the the groups is assumed to be constant with respect its local population density. In reality, the mobility is likely going to be a function of population density owing to inter-population attraction and repulsion. Therefore, highly segregated population groups might exist at a lower Pe than the existing model predicts. Nonetheless, with the simple inclusion of chemotaxis, one can introduce the tendency of clustering and appearance of hotspots. However, chemotaxis alone is not sufficient to capture the complexities of human migration and a heterogeneity in the landscape is also required, which we discuss next.

\begin{figure}[H]
\includegraphics[width=0.45\textwidth]{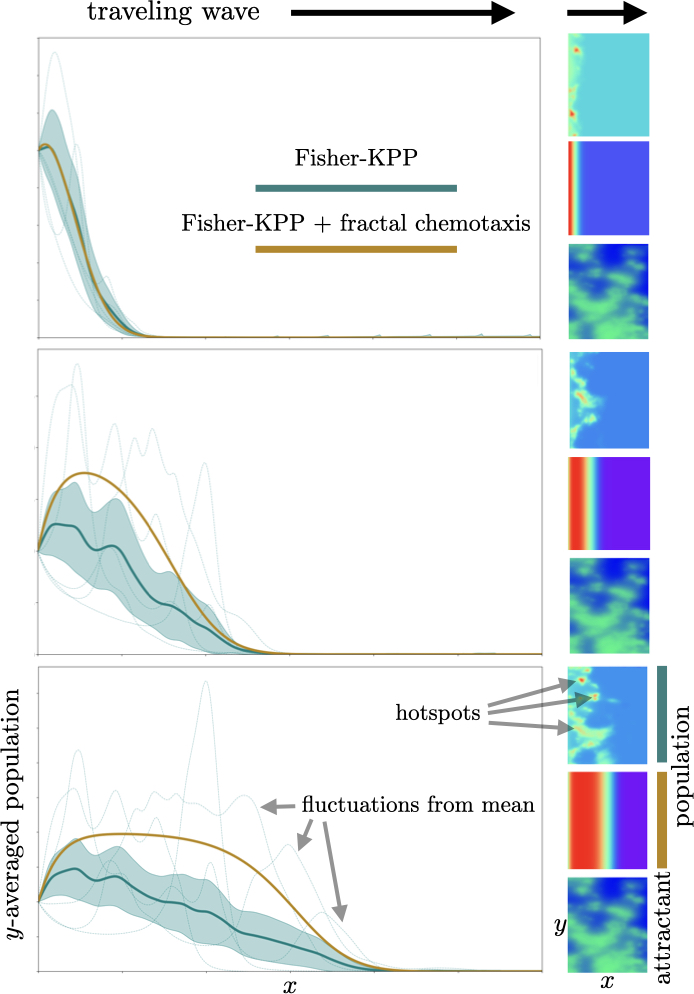}
\caption{\textbf{Advancing wave}. The Fisher-KPP equation permits traveling wave solutions for a reacting and diffusing population. These solutions are characterized by wavefront speed. Including chemotaxis with an attractant distributed heterogeneously over a fractal terrain modifies these traveling wave solutions by introducing heterogeneity in the population as, in motion, it becomes exposed to multi-scale gradients and naturally forms hotspots. Early (top), intermediate (middle), and late (bottom) time steps from a traveling wave solution to equations~\ref{eqn:governingeqs} with and without chemotaxis. Chemotaxis introduces clustering and modifies the population distribution along the wave direction. Because the population growth rate depends on the population itself, there is enhanced growth in the hotspot regions and the total population is distinct from the case without chemotaxis. $\kappa_n=10$, $\kappa_\phi=10$, $r_{n0}=0.1$, $r_{\phi0}=0.1$, $n_\text{max}=2$, $\phi_\text{max}=2$, $\chi=100$.}\label{fig:fisherkpp}
\end{figure}

\section{Traveling wave solution in presence of chemotaxis and fractal terrain}

We now discuss the implications of introducing a chemotaxis with fractal terrain to the classical Fisher-KPP model. First, we focus on the well-known Fisher-KPP scenario, which yields a traveling wave solution. Representative time snapshots are shown in figure~\ref{fig:fisherkpp}, where population density is moving in the $x$-direction. In order to isolate the effect of terrain-dependent chemotaxis, two exemplary cases are shown which retain the essential features of the Fisher-KPP model. In the first (gold line), the standard assumptions of the original Fisher-KPP model are adopted, which in our system is expressed by assigning the chemotactic mobility $\chi=0$ and removing the Gaussian factor from equation~\ref{eqn:popgrowth}. As the population in this case is blind to the terrain, the solution is homogeneous in the $y$-direction. In the second case (teal lines), the chemotactic mobility $\chi>0$ and the population growth term is unchanged. Because of the nonzero chemotactic mobility, the population senses the attractant $\phi$ which must also be computed. For simplicity, $\phi$ is not considered as a traveling wave but disperses throughout the domain immediately. Heterogeneity is introduced by including the Gaussian factor in equation~\ref{eqn:attgrowth} which couples the attractant growth rate to the terrain $h(\boldsymbol x)$, which is chosen to be fractal in this case. This creates heterogeneity in the chemotactic population wave which can be observed in its fluctuations from the $y$-mean. In two-dimensional space, hotspots can be seen in the chemotactic population which are more intense than the relatively shallow gradients of the attractant (an effect of the additional non-linearity in the chemotactic coupling). Furthermore, the nonlinear Fisher-KPP growth term imposes that the two cases, with and without chemotaxis, should not see identical total population because population hotspots experienced disproportionate growth.

Our result is general in that even a very diffuse attractant gradient, over a fractal terrain, can lead to strong clustering. It is natural to expect that including advection in the Fisher-KPP model would alter the traveling wave physics. As  demonstrated in figure~\ref{fig:fisherkpp}, a primary contribution of chemotactic advection when tied to a fractal terrain, is to induce clustering behavior within the traveling wave. Without chemotaxis, the system migrates only due to diffusion and is homogeneous without appreciable fluctuations from the mean. With chemotaxis, the population forms clusters as both small-scale and large-scale attractant gradients guide the population in many directions. Ultimately the population distribution does not directly mimic the attractant distribution but instead balances a complex interplay between growth, diffusion, and advection. The clustering is a natural consequence of introducing multi-scale advection into a system which is otherwise not multi-scale.

\begin{figure*}
\centering
\includegraphics[width=0.95\textwidth]{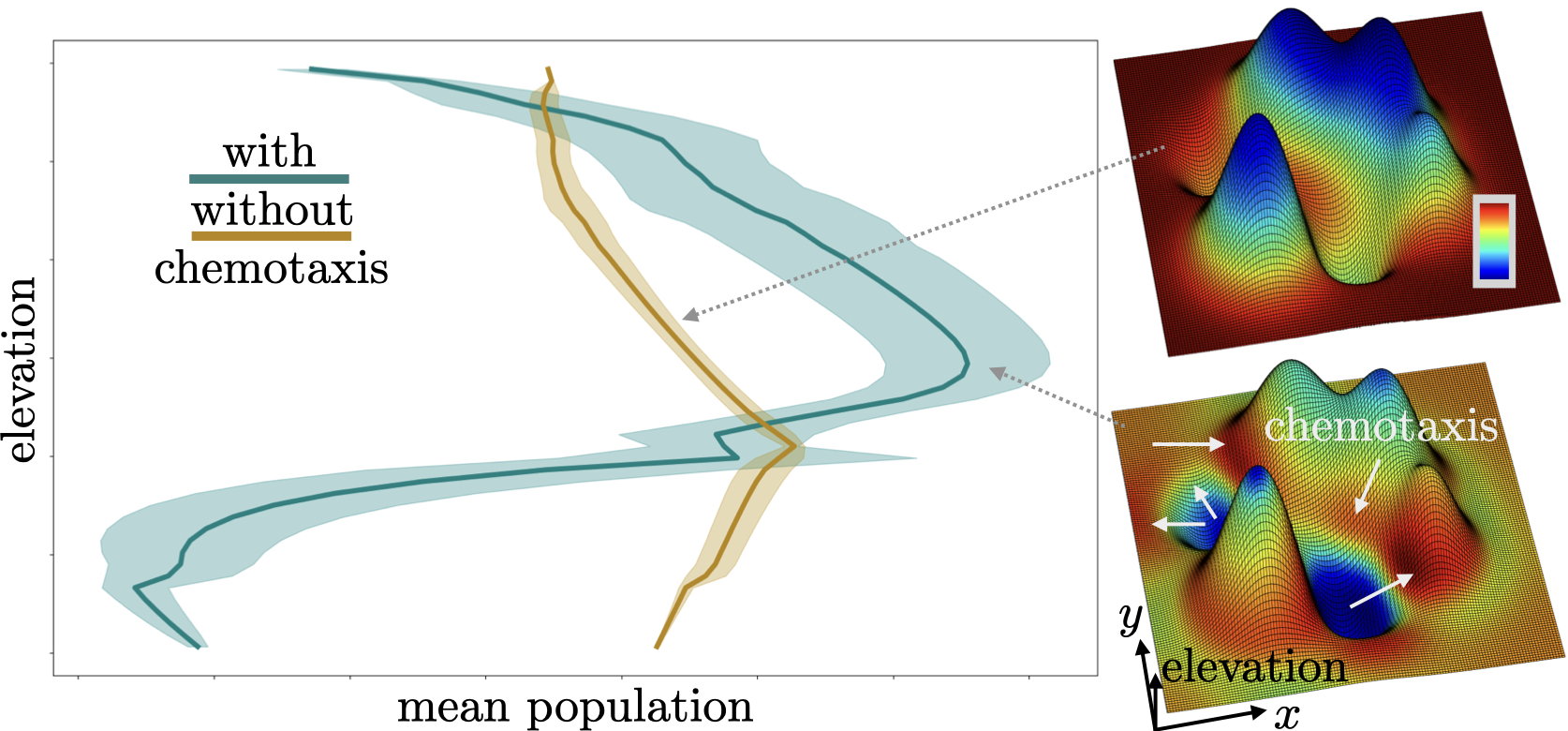}[h]
\caption{\textbf{Migration on a simple heterogeneous terrain}. A population which grows preferentially within a select elevation range may be displaced by the presence of a attractant which grows preferentially in a different elevation range. This holds for even a simple (non-fractal) terrain. The extent to which the population is displaced is identified by comparing the mean elevation with and without chemotaxis. (Left) Spatially-averaged steady-state solution for population from equations~\ref{eqn:governingeqs} along a simple terrain demonstrating chemotaxis-induced heterogeneity. (Right) Spatial distribution of the population biased (top) toward the preferred elevation and (bottom) balanced between preferred elevation and chemotactic migration. Blue to red on the colorscale indicates increasing population. The chemotactic population, due to the enhanced population-dependent growth in the hotspots generated by chemotactic migration, has an increased global concentration. $\kappa_n=80$, $\kappa_\phi=1$, $r_{n0}=0.01$, $r_{\phi0}=0.1$, $n_\text{max}=2$, $\phi_\text{max}=2$, $\chi=100$.}\label{fig:simple}
\end{figure*}

Now we investigate steady-state situations to isolate the effect of chemotaxis on populations which have a preferred growth region. A simple (non-fractal) terrain is chosen and is coupled to the population $n$ through the growth rate (Gaussian factor in equation~\ref{eqn:popgrowth}) in the absence of chemotaxis. The steady-state population distribution is concentrated within the growth envelope owing to a balance between reaction and diffusion (see fig.~\ref{fig:simple} top right). The population can of course be coupled to terrain in a second way, as seen in figure~\ref{fig:fisherkpp}, through chemotaxis along a terrain-dependent attractant. In this system, which now includes a preferred elevation for the population, which is furthermore distinct from the preferred elevation of the attractant, it is apparent that chemotactic advection can influence the population such that it migrates away from its preferred elevation toward regions abundant in the attractant (fig.~\ref{fig:simple} bottom right). In figure~\ref{fig:simple} left, the non-chemotactic population (gold) is maximized at its preferred elevation whereas the chemotactic population (teal) reaches a steady state which, in compromising its desire to be in the proximity of the attractant with its own natural growth rates, finds a new apparent preferred elevation; the population is maximized above its "natural" preferred elevation. For this reason, any large-scale tendencies of how a population is distributed along a heterogeneous landscape cannot necessarily be understood in terms of growth qualities alone. It may be necessary to consider  interactions between the population and the environment, which will consequently create a chemotactic transport that will modify the growth. In a more complicated situation such as the migration of two population groups (say A and B), a chemotactic model may explain the phenomenon of population group B competing with population A if they have a tendency to be attracted to a chemoattractant, despite their differences, as has been reported in migration 
 of skilled and unskilled  labor ~\cite{reichlin1998diverging}.

\begin{figure*}[h]
\centering
\includegraphics[width=0.95\textwidth]{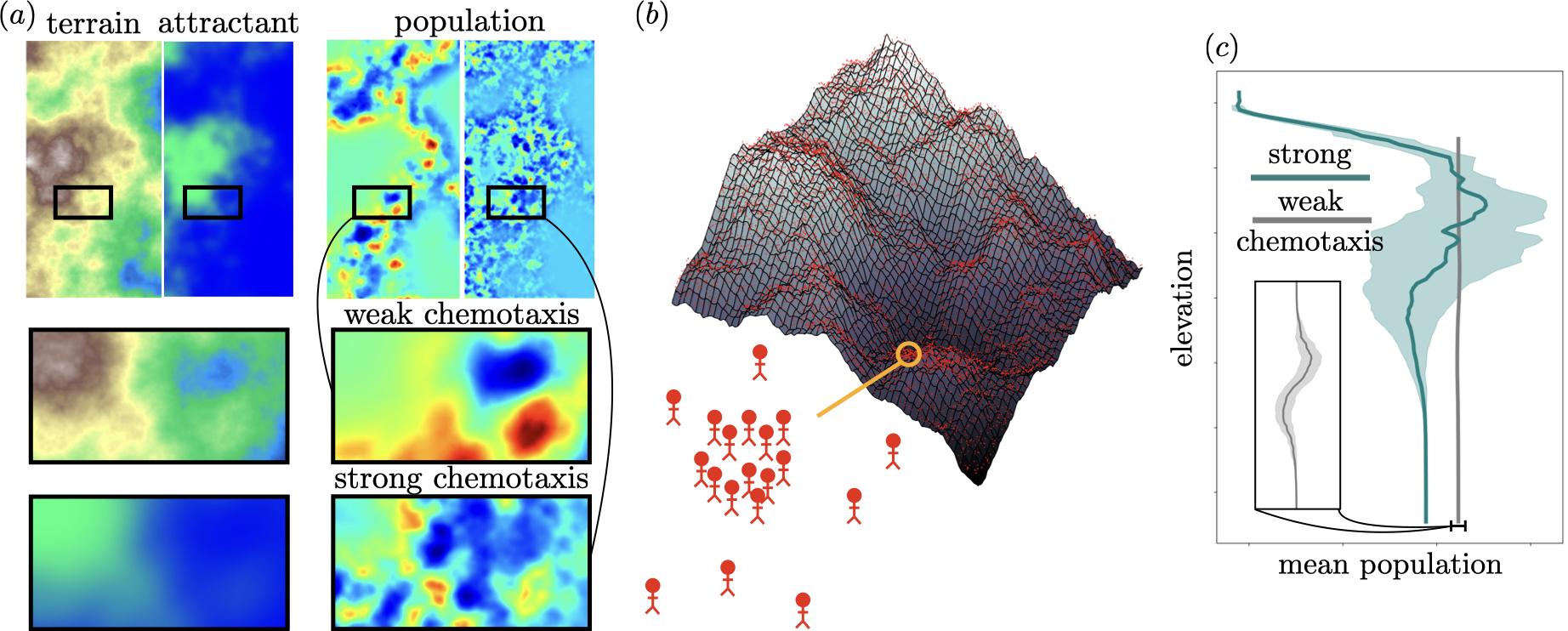}
\caption{\textbf{Chemotactic clustering}. The strength of coupling between the chemotactic population and its attractant has a qualitative effect on the population dispersal. (a) Steady-state solution to equations~\ref{eqn:governingeqs} with fractal terrain $h(\boldsymbol x)$. The degree of population clustering varies with chemotactic mobility. (b) Schematic of population clustering on a multiscale terrain. The terrain may screen a population from global flux balances and lead to a cascade of length scales over which the growth, chemotaxis, and diffusion are balanced. (c) Elevation profiles of population for weak and strong chemotactic mobility. A simple, sinusoidal perturbation to the population which balances mean chemotactic force, growth, and diffusion appears to approximate the mean population distribution versus elevation in the case of weak chemotaxis. However, if the chemotactic mobility is increased, a more complicated elevation profile is established which indicates that the population has formed hotspots which are balanced at a broader range of elevations. Owing to the stronger clustering, the total population growth is enhanced more intensely. $\kappa_n=1$, $\kappa_\phi=10$, $r_{n0}=0.1$, $r_{\phi0}=0.1$, $n_\text{max}=2$, $\phi_\text{max}=2$, $\chi=1$ and $100$.}\label{fig:weakvsstrongsketch}
\end{figure*}

Although it is straightforward to identify how the presence of chemotaxis causes the population to cluster in our system, it is not trivial to understand how the strength of chemotaxis influences the clustering behavior. In figure~\ref{fig:weakvsstrongsketch}, we demonstrate for a fractal terrain how increasing the strength of chemotaxis (through the chemotactic mobility $\chi$) can modify the most prominent clustering length scales. Under weak chemotaxis, the population diffusion is balanced at larger length scales, and larger clusters dominate. Under strong chemotaxis, the population is more confined to the finer length scales of the attractant gradient and approaches the attractant diffusion length scale at which no gradients persist that could drive chemotaxis. Perhaps counter-intuitively, the attractant, which is kept the same in both the weak and strong cases, possesses gradients which are substantially weaker than those of the population in either case, and a rich variety of clustering behavior can be explained by tuning just one parameter (the chemotactic mobility $\chi$) which can be a property of the population independent of its environment. This means that even a small amount of heterogeneity can be sufficient to explain a wide variety of population clustering dynamics, when chemotaxis is considered. A quantitative study of the clustering of a chemotactic population is the subject of future investigation. In figure~\ref{fig:weakvsstrongsketch}(c), we show how the distribution of population with elevation is strongly influenced by chemotaxis and furthermore that the stronger chemotaxis naturally creates strong heterogeneity. In the case of weak chemotaxis (gray), only a subtle gradient in population emerges from the migration which responds most strongly to the shallowest part of the gradient set by the difference in preferred height between the population and the attractant. Under strong chemotaxis, the population disperses over many length scales as it is interacting more strongly with the fractal terrain.

We have demonstrated in figs.~\ref{fig:sharpening}-\ref{fig:weakvsstrongsketch} four key features inherent to our heterogeneous chemotaxis model of population dynamics: intense population increase in response to gradients in attractive qualities, formation of population hotspots which couple to a traveling wave, displacement from inherent tendencies or places of origin, and clustered dispersal which depends qualitatively on the strength of interaction with the environment. Our model may be a useful starting point to explain the complicated dynamics of human migration throughout history. As pointed out by Cohen~\cite{cohen1992nonlinearity}, chemotactic behavior may explain such mysteries as the origin of the Basque language. We might also expect that other ways in which populations become distinct, such as political polarization or class hierarchy, can come from chemotactic social interactions on a fractal landscape of ideas and identity, which may fit broadly into the mathematical framework of chemotaxis. As this point, we would also like to clarify that we do not intend to suggest that the model described above can capture all the complexities of human population dynamics. However, we underscore that chemotaxis is an essential transport process that has been overlooked in the human dispersal analyses.

\section{Outlook}
In the study we have identified chemotactic migration on fractal terrains as a key mechanism for population clustering. This synthesizes a small body of literature which recognized in a human migration context various elements of the Keller-Segel framework used overwhelmingly only for microbiological population dispersal modeling~\cite{cohen1992nonlinearity,gu2017stationary}, connection between advective processes and terrain for human migrations~\cite{timmermann2016late,gu2017stationary}, and the importance of the fractal character of the terrain for human migration~\cite{johnson1992animal,hamilton2007spatial}. By combining these concepts and drawing analogy to soft matter systems, we find a simple way to explain the ubiquitous clustering of human populations.

More work is needed to understand the various ecological, economical, or cultural contributions to chemoattractants throughout all stages of human history. Furthermore, our framework will serve as a starting point for researchers studying dynamical, systematic interactions between human populations and the environment, which will be especially relevant for understanding future climate scenarios based off of physical models~\cite{masson2022global}. Various physical effects such as time lags~\cite{fort1999time} and fractal diffusion~\cite{hamilton2007spatial} will be considered in future models. For modern population dynamics, non-local transport phenomena such as airplane travel may be included \textit{via} integral equations. This might enhance clustering in a way that is complementary to chemotactic advection. Although human migration data is sparse, analogous systems can be designed in soft matter experiments from which physical understandings of human migration can emerge. Such effects as complex geometries and confinement which have been well-studied in microbiological situations~\cite{martinez2022morphological} may prove highly relevant at the human scale.

\par{} This article expands the capability of human population models by including chemotactic advection. The proposed framework is particularly relevant because human migration patterns are likely to become increasingly important due to the impact of climate change~\cite{masson2022global,bartusek20222021}. In particular, chemotactic movement of humans due to concentration gradients in safety indices may play a dominant role. Additionally, this methodology could be used to study the segregation of different social groups in cities, the formation of social networks, and the spread of cultural trends. The phenomenon of chemotaxis could also be included in agent-based population models, which are microsimulation schemes that allow for more complex interactions between individuals. Agent-based modeling of human population dynamics for situtations such as disease transmission~\cite{kailasham2024influence} may be crucial for designing realistic continuum models to go beyond the idealized equations and parameters presented in this manuscript. Finally, soft matter physicists can take inspiration and consider designing models and experiments that directly impact anthropological models.

\section*{Acknowledgements}
A.G. thanks the National Science Foundation (CBET - 2238412) CAREER award. B.M.A. thanks the National Science Foundation for the financial support through the Graduate Research Fellowship, and Stanford University for the Stanford Graduate Fellowship.

\section*{Author Contributions}
B.M.A. and A.G. conceptualized the project. B.M.A. wrote the simulation codes. B.M.A. and A.G. analyzed the results, conducted the literature review, developed the mathematical models and wrote the manuscript. 

\bibliography{rsc} 
\bibliographystyle{rsc} 

\clearpage

\end{document}